\documentclass[oldversion]{aa}

\usepackage{graphicx}

\usepackage{txfonts}

\usepackage{color}

\usepackage{natbib}

\usepackage{color}

\newcommand{\qm}[1]{``#1''}
\newcommand{\fromto}{\,${--}$\,}

\def\sss{\scriptscriptstyle}
\def\U{{\sss \!U}}
\def\L{{\sss \!L}}
\def\K{{\sss \!K}}

\def\nur{\nu_\mathrm{r}}
\def\nuv{\nu_\theta}
\def\nuL{\nu_\L}
\def\nuU{\nu_\U}
\def\nuK{\nu_\K}

\def\factor{\mathcal{F}}

\definecolor{gray}{rgb}{.6,.6,.6}
\definecolor{green}{rgb}{0,.6,0}
\definecolor{red}{rgb}{.9,0,0}
\newcommand*{\change}[1]{{{#1}}}

\begin{document}

\title{Disc-oscillation resonance and neutron star QPOs: \\3:2 epicyclic orbital model}

\author
{Martin Urbanec$^1$, Gabriel T\"or\"ok$^1$, Eva \v{S}r\'amkov\'a$^1$, Petr \v{C}ech$^1$, Zden\v{e}k Stuchl\'{\i}k$^1$, Pavel Bakala$^1$
}


\institute{$^1$ Institute of Physics, Faculty of Philosophy and Science, Silesian
  University in Opava, Bezru\v{c}ovo n\'{a}m. 13, CZ-74601 Opava, Czech Republic
  }

\date{Received / Accepted}

\authorrunning{M. Urbanec et al.}
\titlerunning{On 3:2 epicyclic resonance in NS kHz QPOs}

\date{Accepted 23/06/2010.}

\abstract
{The high-frequency quasi-periodic oscillations (HF QPOs) that appear in the X-ray fluxes of low-mass X-ray binaries remain an unexplained phenomenon. Among other ideas, it has been suggested that a non-linear resonance between two oscillation modes in an accretion disc orbiting either a black hole or a neutron star plays a role in exciting the observed modulation. Several possible resonances have been discussed.  A particular model assumes resonances in which the disc-oscillation modes have the eigenfrequencies equal to the radial and vertical epicyclic frequencies of geodesic orbital motion. This model has been discussed for black hole microquasar sources as well as for a group of neutron star sources. Assuming several neutron (strange) star equations of state and Hartle-Thorne geometry of rotating stars, we briefly compare the frequencies expected from the model to those observed. Our comparison implies that the inferred neutron star radius $R_{\mathrm{NS}}$ is larger than the related radius of the marginally stable circular orbit $r_{\mathrm{ms}}$ for nuclear matter equations of state and spin frequencies up to $800$Hz. For the same range of spin and a strange star (MIT) equation of state, the inferrred radius is $R_{\mathrm{NS}}\sim r_{\mathrm{ms}}$. The "Paczy\'nski modulation" mechanism considered within the model requires that $R_{\mathrm{NS}}<r_{\mathrm{ms}}$. However, we find this condition to be fulfilled only for the strange matter equation of state, masses below $1M_{\sun}$, and spin frequencies above 800Hz. This result most likely falsifies the postulation of the neutron star 3:2 resonant eigenfrequencies being equal to the frequencies of geodesic radial and vertical epicyclic modes. We suggest that the 3:2 epicyclic modes could stay among the possible choices only if a fairly non-geodesic accretion flow is assumed, or if a different modulation mechanism operates.}

\maketitle

\section{Introduction: HF QPOs and desire for strong-gravity}
\label{section:introduction}

Galactic low mass X-ray binaries (LMXBs) display \emph{quasi-periodic oscillations (QPOs)} in their observed X-ray fluxes (i.e., peaks in the X-ray power density spectra). Characteristic frequencies of these QPOs range from \mbox{$\sim\!10^{-2}$\,Hz} to \mbox{$\sim\!10^{3}$\,Hz}. Of particular interest are the so-called high-frequency (HF) QPOs with frequencies  typically in the range $50\fromto1300\,\mathrm{Hz}$, which is roughly of the same order as the range of frequencies characteristic of orbital motion close to a low mass compact object. We briefly recall that there is a crucial difference between HF QPOs observed in black hole (BH) and neutron star (NS) systems. {In BH systems, the HF QPO peaks are commonly detected at constant (or nearly constant) frequencies that are characteristic of a given source.} When two or more QPO frequencies are detected, they usually come in small-number ratios, typically in a \mbox{$3\!:\!2$} ratio \citep[][]{abr-klu:2001,klu-abr:2001, mc-rem:2003,tor-etal:2005}. For NS sources, on the other hand, HF (or kHz) QPOs often appear as \emph{twin QPOs}. These features, on which we focus here, consist of two simultaneously observed peaks with distinct actual frequencies that substantially change over time. 
The two peaks forming twin QPO are then referred to as \emph{the lower and upper QPO} in agreement with the inequality in their frequencies.

The amplitudes of twin QPOs in NS sources are typically much stronger and their coherence times much higher than those in BH sources \citep[e.g.][]{mc-rem:2003,bar-etal:2005a,bar-etal:2005b,bar-etal:2006,men:2006}. It is however interesting that most of the twin QPOs with high statistical significance have been detected at lower QPO frequencies $600\fromto700$Hz and upper QPO frequencies $900\fromto1200$Hz. Because of this the twin QPO frequency ratio clusters mostly around $\approx$\,3:2 value posing thus some analogy to BH case \citep[see][for details and a related discussion]{abr-etal:2003a, bel-etal:2007b, tor-etal:2008a, tor-etal:2008b, tor-etal:2008c, bou-etal:2009}. 
In several NS sources, the difference in the amplitudes of the two peaks changes sign as their frequency ratio passes through the (same) 3:2 value \citep{tor:2009}. A detailed review on the other similarities and differences in the HF QPOs features can be found in \cite{Kli:2006:CompStelX-Ray:}. 

\subsection{HF QPO interpretation}

There is strong evidence supporting the origin of the twin QPOs inside 100 gravitational radii, $r_{\mathrm{g}}$\,=\,\mbox{G$M$c$^{-2}$}, around the accreting compact objects \cite[e.g.,][]{Kli:2006:CompStelX-Ray:}. At present, there is no commonly accepted QPO theory. It is even unclear whether this theory could involve the same phenomena for both BH and NS sources. Several models have been proposed to explain the HF QPOs, most of which involve orbital motion in the inner regions of an accretion disc. When describing the orbital motion, the Newtonian approach necessarily fails close to the compact object. Two of the most striking differences arise from the relevant general relativistic description: Einstein's strong gravity cancels the equality between the Keplerian and epicyclic frequencies, and (due to the existence of the marginally stable circular orbit $r_{\mathrm{ms}}$) it applies a limit to the maximal allowed orbital frequency. Several effects such as the relativistic precessions of orbits then pop up in the inner accretion region. Finding a proper QPO model may thus help us to test the strong field regime predictions of general relativity and, in the case of NS sources, also the models of highly dense matter  \citep[see][for a review]{Kli:2006:CompStelX-Ray:,Lam-Bou:2007:ASSL:ShrtPerBS}.

\subsection{Non-linear resonances between \qm{geodesic and non-geodesic} disc-oscillations}

Numerous explanations of the observed lower and upper HF modulation of the X-ray flux have been proposed while hypothetical resonances between the two QPO oscillatory modes are often assumed. Specific ideas considering \emph{non-linear resonances between disc-oscillation modes} have been introduced and extensively investigated by Abramowicz, Klu{\'z}niak and collaborators \citep[][and others; see also \citeauthor{ali-gal:1981}, 1981 and \citeauthor{ali:2006}, 2007]{klu-abr:2001,abr-klu:2001,abr-etal:2003a,abr-etal:2003b,reb:2004,tor-etal:2005,hor:2008,stu-etal:2008b,hor-etal:2009}. These ideas have been widely discussed and adapted into numerous individual disc-oscillation models.

The subject of disc oscillations and their propagation has been extensively studied analytically for thin disc (i.e., nearly geodesic, radiatively efficient) configurations \citep[][]{oka-etal:1987,kat-etal:1998,wag:1999,wag-etal:2001,sil-etal:2001,ort-etal:2002,wag:2008}. The derived results  have been compared to those for \qm{thick} (radiatively inefficient, slim-disc or toroidal) configurations foe which both analytical \citep{bla:1985,sra:2005,abr-etal:2006,bla-etal:2006,bla-etal:2007,str-sra:2009} and numerical \citep{rez-etal:2003a,rez-etal:2003b,rez:2004,mon-etal:2004,zan-etal:2005,sra-etal:2007} studies have been performed. Several consequences of disc-oscillation QPO models have been sketched, some having direct relevance to non-linear resonance hypotheses. In particular, it has been found that, due to pressure effects, the values of the frequencies at radii fixed by a certain frequency ratio condition can differ between the geodesic and fairly non-geodesic flow of factors such as 15$\%$ \citep[][]{bla-etal:2007}.

\subsection{Aims and scope of this paper}

\citet{gon-klu:2002} suggested that the resonance theory of kHz QPOs can help us to discrimine between quark (strange matter) stars and neutron stars. In the spirit of this suggestion, we examine a particular, often quoted  \qm{3:2 epicyclic resonance model} (or rather a class of these models). The paper is arranged as follows.

In Sect.~\ref{section:model}, we briefly highlight some important aspects of non-linear resonance models specific to neutron stars and a 3:2 epicyclic resonance model. In Sect.~\ref{section:MR}, we compare the model to the HF QPO observations of a group of NS sources displaying the 3:2 ratio. The restrictions to the mass and radius implied by the equations of state for non-rotating NS are included. In Sect.~\ref{section:spin}, we explore the corrections required for NS rotation and again consider the equations of state. In Sect.~\ref{section:conclusions}, we assign some consequences and discuss possible falsification of the model, whereas the nearly geodesic and fairly non-geodesic cases are considered separately.

Throughout the paper, we use the standard notation where $\nuL,~\nuU$ represent the observed lower and upper QPO frequencies, while $\nuK,~\nur,~\nuv$ represent the Keplerian, radial epicyclic and vertical epicyclic frequencies for the considered spacetime and its parameters.

\section{Resonances in discs around neutron stars}
\label{section:model}

 Miscellaneous variations in the non-linear, disc-oscillation resonances have been discussed in the past \citep[see, e.g.,][]{abr-klu:2001, abr-klu:2004, tor:2005a, hor-kar:2006}. While the basic approaches have been common to both black-hole and neutron-star models, several differences between the two classes of sources have been considered. In particular, it has been suggested that, in a turbulent NS accretion flow, the resonant eigenfrequencies are not fixed \cite[e.g., when oscillations of a tori changing its position are assumed;][]{zan-etal:2003,rub-lee:2005,abr-etal:2006,tor-etal:2007,klu-etal:2007}, or that the resonant corrections to eigenfrequencies reach high values \citep{abr-etal:2003b,abr-etal:2005a,abr-etal:2005b}. Both possibilities are taken into account in Sect.~\ref{section:MR}.

\subsection{Modulation}

One more important difference between the two aforementioned source types concerns the QPO modulation mechanism \citep{bur-etal:2004,hor:2005a,abr-etal:2007,bur:2008}. In the black hole case, the weak modulation is assumed to be primarily connected to radiation of the oscillating disc and the related relativistic lensing, light-bending, and Doppler effects. In the neutron star case, the expected modulation is connected to the flux emitted from a hot spot on the NS surface causing a strong QPO amplitude.

\begin{figure}[t!]
\begin{center}
\begin{minipage}{1\hsize}
\hfill
\includegraphics[width=0.95\textwidth]{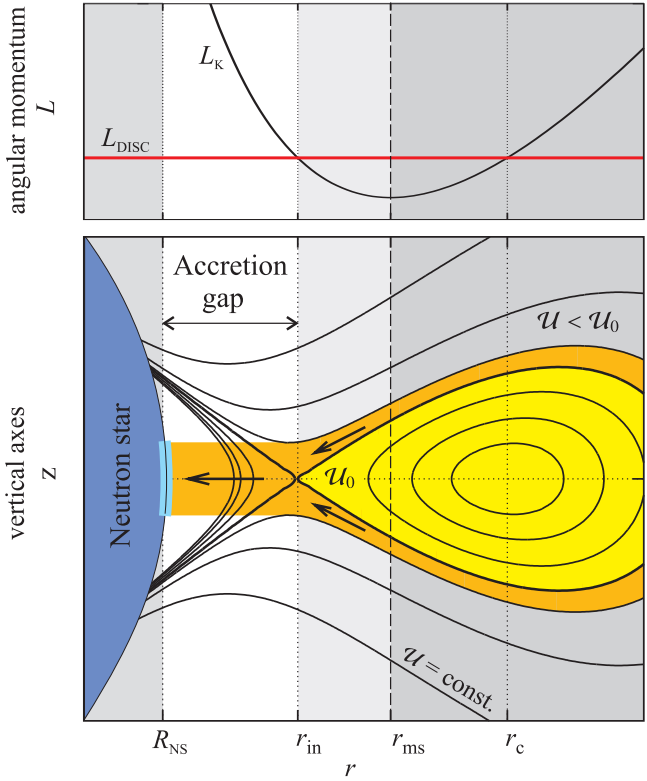}
\end{minipage}
\end{center}
\caption{\emph{Mass-flow leaving the disc and crossing the relativistic accretion gap \citep[after][]{abr-etal:2007}.} Top: Keplerian angular momentum versus the angular momentum in the flow. Bottom: The equipotential surfaces and the distribution of fluid in a meridional cross-section of the disc-configuration. The yellow area denotes the fluid in the disc, while the orange area corresponds to the overflow modulated by the oscillations. Enhanced luminosity arises as the flow enters the boundary layer (light-blue colour).}
\label{figure:gap}
\end{figure}

 We briefly describe the \emph{\qm{Paczy\'nski-modulation} mechanism} \citep{pac:1987}, which was investigated by \citet{hor:2005a} and \citet{abr-etal:2007}. The schematic Fig.~\ref{figure:gap} displays the considered situation. The expected mass flow is described by the Bernoulli equation, while surfaces of constant enthalpy, pressure, and density coincide with surfaces of constant effective potential $\mathcal{U}(r, z)=$~constant \citep[][]{abr:1971}.  The disc equilibrium can exist if the disc surface corresponds to one of the equipotentials inside the so-called Roche lobe (region indicated by the yellow colour). No equilibrium is possible in the region of $r<r_\mathrm{in}$. For a given accretion rate the dynamical mass loss occures when the fluid distribution overflows the surface of the disc for $\mathcal{U}_0=\mathcal{U}(r_{\mathrm{in}})$. When the accretion disc oscillates, it slightly changes its position with respect to the equipotencial surfaces. At a particular location corresponding to the crossing of the equipotentials, the so-called cusp, even a small displacement of the disc causes a large change in the accretion rate. The change in the accretion rate is then nearly instantly reflected by the hot-spot temperature leading to an enhanced X-ray emission \citep{pac:1987, hor:2005a, abr-etal:2007}.

The existence of the surface $\mathcal{U}_0$ above the neutron star is crucial to the model. Therefore, as a necessary condition for its applicability, it is required that
\begin{equation}
\label{equation:modulation-condition}
R_{\mathrm{NS}}/r_{\mathrm{ms}}<1,
\end{equation}
where $R_{\mathrm{NS}}$ denotes the neutron (strange) star radius \citep[\qm{accretion gap paradigm},][]{klu-wag:1985,klu-etal:1990}. We note that this is a necessary but insufficient condition, since the inner radius $r_{\mathrm{in}}$ is located between the marginally stable and marginally bound circular orbit \citep{koz-etal:1978}.

\subsection{Epicyclic resonance}

A particular example of the non-linear resonance between disc-oscillation modes is represented by the concept of the \qm{3:2 epicyclic internal resonance}. This hypothesis is widely discussed \citep[e.g.,][among the other references in this paper]{abr-etal:2002,klu-abr:2002,klu-abr:2005,hor:2004,hor:2005b,kli:2005,tor-stu:2005,vio-etal:2006,reb:2008,rey-mill:2009}. It assumes that the resonant modes have eigenfrequencies equal to radial and vertical epicyclic frequency of geodesic orbital motion given by
\begin{equation}
\nuL^0=\nur(r_{3:2}), \quad \nuU^0=\nuv(r_{3:2}),
\end{equation}
associated with the orbital radius $r_{3:2}$, where $\nuv/\nur=3/2$. We emphasize that models consider oscillations of fluid configurations rather than test particle motion \citep[see, e.g.,][for some details and related references]{klu:2008}.
In the following sections we consider the 3:2 epicyclic model and  whether the Paczy\'nski modulation mechanism may be at work.

\section{NS mass and radius implied by the 3:2 epicyclic resonant model}
\label{section:MR}

In resonance models of BH QPOs, the observed constant frequencies are expected to coincide with the resonant eigenfrequencies. Assuming a particular resonance, one may then relate the black hole spin or mass to the observed frequencies. This procedure was followed by \cite{abr-klu:2001} and later by \cite{tor-etal:2005} and \cite{tor:2005b} for various resonances and sets of sources. In principle, similar calculations can also be made for resonance models of NS QPOs. For neutron stars, the observed frequencies, however, change over time and, moreover, monotonic positive frequency correlations are similar, but specific to the individual sources. Within the framework of the resonance models we can consider two distinct \emph{simplifications} to the observed frequency correlations when inferring the neutron star mass:
\begin{itemize}
\item[a)] The observed frequencies are roughly equal to the resonant eigenfrequencies and the observed frequency correlation follows from the changes in eigenfrequencies
\begin{eqnarray}
\label{equation:a}
\nuL=\nuL^0(r_{3:2}+\Delta r),\quad
\nuU=\nuU^0(r_{3:2}+\Delta r),
\end{eqnarray}
implying~for~the~3:2~epicyclic~model~that
\begin{eqnarray}
\nuL=\nur(r_{3:2}+\Delta r),\quad
\nuU=\nuv(r_{3:2}+\Delta r).
\label{equation:a:epicyclic}
\end{eqnarray}
\item[b)] The eigenfrequencies are constant and the observed correlation is caused by the resonant corrections
\begin{eqnarray}
\nuL=\nuL^0 + \Delta\nuL,\quad
\nuU=\nuU^0 + \Delta\nuU,
\label{equation:b}
\end{eqnarray}
implying~for~the~3:2~epicyclic~model~that
\begin{eqnarray}
\nuL=\nur(r_{3:2})+\Delta\nuL,\quad
\nuU=\nuv(r_{3:2})+\Delta\nuU.
\label{equation:b:epicyclic}
\end{eqnarray}
\end{itemize}
We note that for a) the resonance plays rather a secondary role in producing QPOs, while for b) it represents their generic mechanism.

\subsection{Mass}
\label{section:mass}

In this section, we neglect the effects of neutron star spin and assume the 3:2 epicyclic resonance model in the Schwarzschild spacetime.\footnote{We consider here this standard spacetime description for non-rotating neutron stars, although some alternatives have been discussed in a similar context \citep[see][]{kot-etal:2008,stu-kot:2009}.} Introducing a relativistic factor \mbox{$\factor\equiv c^3/(2\pi GM)$}, Eq.~(\ref{equation:a:epicyclic}) reads
\begin{eqnarray}
\nonumber
\nuU&=&\nuv = \nuK \equiv {x^{-3/2}}\,\factor,\quad x\equiv r/M,\\
\nuL&=&\nur \equiv \nuK\,\sqrt{1-\frac{6}{x}},
\label{equation:frequencies}
\end{eqnarray}
implying~that
\begin{equation}
\nuL=\nuU\sqrt{1-6 \left(\frac{\nuU}{\factor}\right)^{2/3}}. \label{equation:result:a}
\end{equation}
It has been previously discussed in terms of a correlation between the QPO frequency ($\nuL$ or $\nuU$) and frequency difference $\Delta\nu=\nuU-\nuL$ that the correlation given by Eq.~(\ref{equation:result:a}) clearly disagrees with the observations of NS sources \citep[e.g.,][]{bel-etal:2005}. The 3:2 epicyclic resonance model that is fully based on Eq.~(\ref{equation:a}) is therefore excluded. Hence, in the following we focus on the option represented by Eq.~(\ref{equation:b}).

The relation of Eq.~(\ref{equation:b}) to the observation of several NS sources was considered by \cite{abr-etal:2005a,abr-etal:2005b}. 
They assumed that the corrections $\Delta \nu$ in Eq. (\ref{equation:b:epicyclic}) vanish when the observed frequency ratio $\nuU/\nuL$ reaches the 3/2 value. They suggested that the resonant eigenfrequencies $[\nuL^0,~\nuU^0]$  in a group of twelve NS sources are roughly equal to  $[600\mathrm{Hz},~\mathrm{900Hz}]$. For the 3:2 epicyclic model we then find that
\begin{equation}
\label{equation:600:900}
\nur^{3:2} = 600\mathrm{Hz}, \quad \nuv^{3:2}= 900\mathrm{Hz},
\end{equation}
which, from terms given in equation (\ref{equation:frequencies}), implies that the relevant mass must be  around
 $M=1M_{\sun}$ (as first noticed by Bursa~2004 unpublished).

\subsection{Radius}
\label{section:radius}

\begin{figure}[t!]
\begin{minipage}{1\hsize}
\includegraphics[width=.975\textwidth]{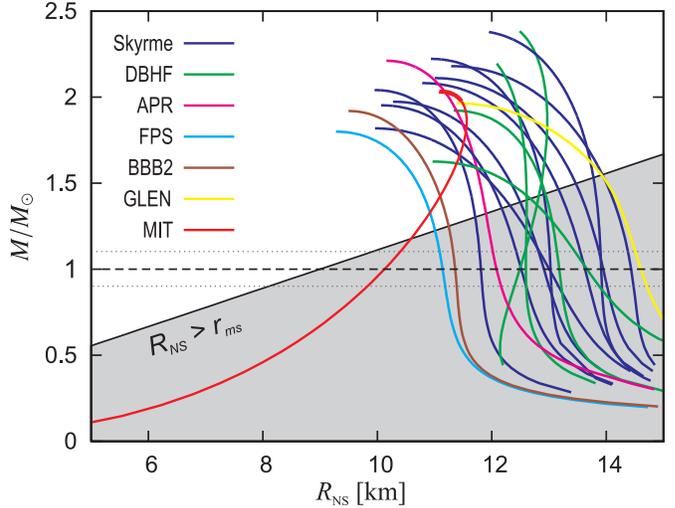}
\end{minipage}
\caption{Mass-radius relations for several EoS assuming a non-rotating star. The shadow area indicates the region with NS radii higher than the radius of the marginally stable circular orbit (no accretion gap). The mass $M=1\pm0.1M_{\sun}$ is denoted by the dashed and dotted horizontal lines.}\label{figure:EoS}
\end{figure}

Modelling of NS equations of state (EoS) have been extensively
developed by numerous published methods and codes (see, \citeauthor{lat-pra:2001},
\citeyear{lat-pra:2001} and \citeauthor{lat-pra:2007},
\citeyear{lat-pra:2007} for a review). Here we calculate NS radii
following  the approach of \citet{har:1967}, \citet{har-tho:1968},
\citet{cha-mil:1974}, and \citet{mil:1977}.  In Fig.~\ref{figure:EoS}, we plot the mass-radius relations for
several EoS.

{\it Skyrme} represents nine  different EoS (namely SkT5, SkO',
SkO, SLy4, Gs, SkI2, SkI5, SGI, and SV) given by the different parameterizations of the effective Skyrme potentials 
\citep[see][and references therein]{rik-etal:2003}. {\it DBHF} represents four different parameterizations, chosen to describe
matter in the framework of Dirac-Brueckner-Hartree-Fock theory.
In particular, we choose the parameterizations labeled HA, HB, LA, and MA
in \citet{kot-gmu:2003} used by \citet{urb-etal:2010} to describe
the properties of static neutron stars. The EoS labeled {\it APR} has often been used. We chose the model labeled $A18
+ \delta v + UIX*$ in the original paper \citep{akm-etal:1998}. Remaining pure neutron-star equations of state are {\it FPS}
\citep{pan-rav:1989} and {\it BBB2} \citep{bal-etal:1997}. The model
labeled {\it GLENDNH3} also includes hyperons \citep{gle:1985}.

The {\it MIT} model represents strange stars calculated using the so-called MIT bag model \citep{cho-etal:1974}, where we used the standard values $B=10^{14} \,\mathrm{g.cm}^{-3}$ for the bag constant and $\alpha_\mathrm{c} = 0.15$ for strong interaction coupling constant.

From Fig.~\ref{figure:EoS} we can see that for $M\sim1M_{\sun}$ the NS radii are in all cases above $r_\mathrm{ms}$. Thus, assuming the Schwarzschild metric the condition presented in Eq.~(\ref{equation:modulation-condition}) is not fulfilled for the X-ray modulation given by the 3:2 epicyclic model.

\section{Effects related to NS spin}
\label{section:spin}

We have restricted attention to the implications of Eq.~(\ref{equation:600:900}) for non-rotating NS. The spin of the astrophysical compact objects and related oblateness however introduce some modifications of the Schwarzschild spacetime geometry. Without the inclusion of magnetic field effects, it has been found that the rotating spacetimes induced by most of the up-to-date neutron star equations of state (EoS) are well approximated with the solution of \citet{har-tho:1968} \citep[see][for details]{ber-etal:2005}. We use this solution (in next HT) to discuss the spin corrections to the above results.

The HT solution reflects three parameters, the neutron star mass $M$, angular momentum $J$, and quadrupole moment $Q$.  We note that the Kerr geometry represents the \qm{limit} to the HT geometry for $\tilde{q}\equiv QM/J^2\rightarrow1$ up to the second order in $J$. The formulae for Keplerian and epicyclic frequencies in the HT spacetime were derived by \citet{abr-etal:2003c}. We applied these formulae to solve Eq.~(\ref{equation:600:900}). Figure~\ref{figure:MJ}  displays the resulting surface colour-scaled in terms of $M/M_{\sun}$, $j=cJ/GM^2$, and $\tilde{q}$. We can see that for low values of $\tilde{q}$ and any $j$ the implied $M$ increases with increasing $j$, while exactly the opposite dependence $M(j)$ occurs for high values of $\tilde{q}$ and $j\geq0.2$.

\subsection{EoS and radii}

\begin{figure}[t!]
\begin{minipage}{1\hsize}
\vspace{1.2ex}

\begin{center}
\includegraphics[width=1\textwidth]{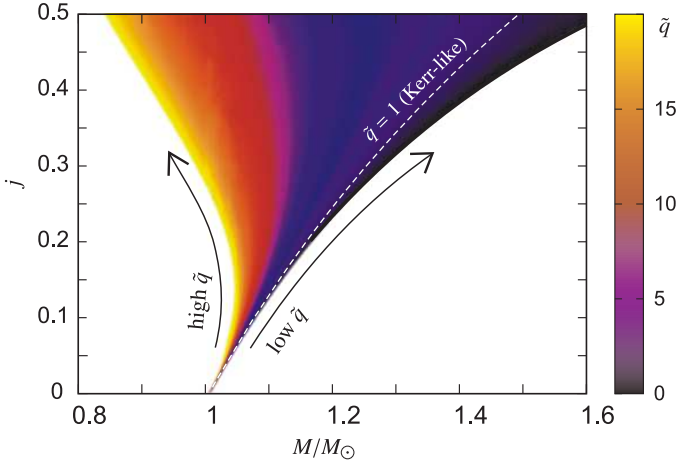}
\end{center}
\end{minipage}
\caption{Solution of the 3:2 frequency equation (\ref{equation:600:900}) projected onto $M-j$ plane and colour-scaled in terms of $\tilde{q}$.}\label{figure:MJ}
\end{figure}

For a given EoS, the parameter $\tilde{q}$ decreases with increasing $M/M_\mathrm{max}$. In more detail, it is usually $\tilde{q}\sim10$ for $1M_{\sun}$, while $\tilde{q}\in(1.5,~3$) for the maximal allowed mass \citep[e.g.][Fig.~3 in their paper]{tor-etal:2010}. Since the non-rotating mass inferred from the model is about $1M_{\odot}$, one can expect that realistic NS configurations will be related to $M\fromto j$ solutions associated with high $\tilde{q}$. These are denoted in  Fig.~\ref{figure:MJ} by colours of the yellow-red spectrum.

We checked this expectation using the same set of EoS as in Sect.~\ref{section:radius}. We calculated the configurations for each EoS covering the range of the central density $\rho_\mathrm{c}$ implying that $M\in(\approx0.5M_{\sun},~M_{\mathrm{max}})$ and the spin frequency $\nu_\mathrm{s}\in(0,~\nu_\mathrm{max})$, using thousand bins in each of both independent quantities. The mass $M_{\mathrm{max}}$ is the maximal mass allowed for a given EoS and spin frequency $\nu_\mathrm{s}$. The frequency $\nu_\mathrm{max}$ is the maximal frequency of a given neutron star and is equal to the Keplerian frequency at the surface of the neutron star at the equator, corresponding to the so-called mass-shedding limit. In this way, we obtained a group of $15\times1000^2\approx10^7$ configurations. From these we retained only those fulfilling the condition in Eq.~(\ref{equation:600:900}) for the epicyclic frequencies \citep[$\nu\!=\!\nu\,(M,\,j,\,q),~$][]{abr-etal:2003c}
 extended to 
\begin{equation}
\label{equation:600:900:expanded}
2/3\nuv^{3:2}=\nur^{3:2}\in~(580\mathrm{Hz},~680\mathrm{Hz}).
\end{equation}
We note that this range of considered eigenfrequencies is based roughly on the range of the observed 3:2 frequencies \citep{abr-etal:2005a,abr-etal:2005b}. The combinations of mass and angular momentum selected in this way are displayed in Fig.~\ref{figure:MJEoS}a. Inspecting the figure, one can see that the mass decreases with increasing $j$ above $j\sim0.3$.

Figure~\ref{figure:MJEoS}b indicates the ratio $R_\mathrm{NS}/r_\mathrm{ms}$ for the selected configurations (shown in Fig.~\ref{figure:MJEoS}a). The modulation-condition presented in Eq.~(\ref{equation:modulation-condition}) appears only to be fulfilled for MIT-EoS and high spins above 800Hz.

\section{Discussion and conclusions}
\label{section:conclusions}

\begin{figure}[t!]
~\hfill
\begin{minipage}{.96\hsize}
\noindent
a)\\
\includegraphics[width=1\textwidth]{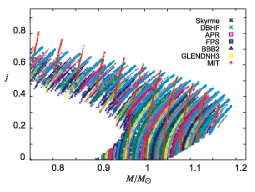}\\\\\\
b)\\
\includegraphics[width=1\textwidth]{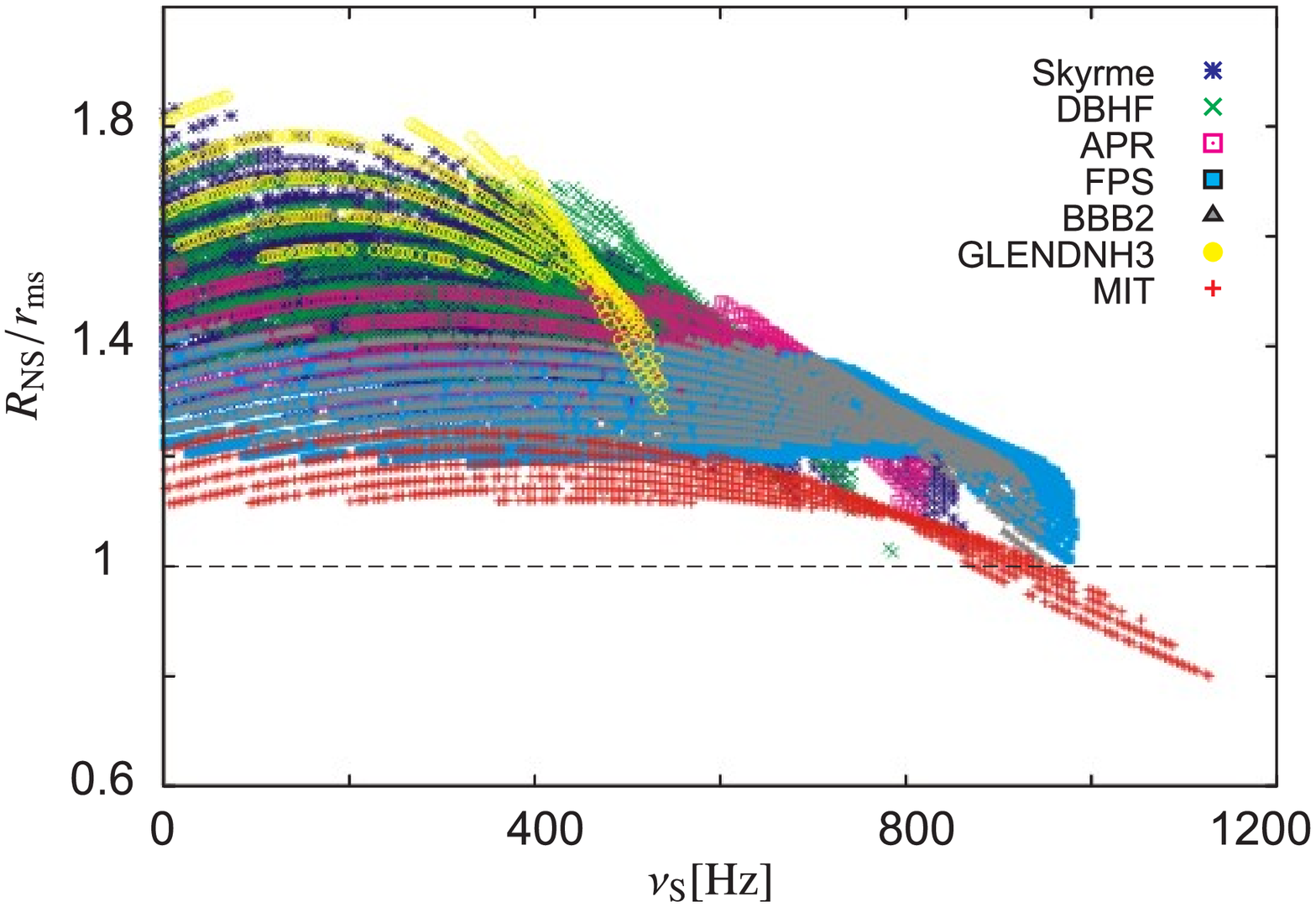}
\end{minipage}
\caption{{\bf{a)}}  NS configurations fulfilling the 3:2 frequency
condition (\ref{equation:600:900:expanded}). {\bf{b)}} Related
relationships between NS spin and radii evaluated in terms of ISCO
radii $r_\mathrm{ms}(M,~j,~q)$. Only subset of
configurations obeying the condition
(\ref{equation:600:900:expanded}) are depicted in the figures for
clearness. Lines tending to appear on both figures correspond to
configurations with same central parameters and different rotational
spin. } \label{figure:MJEoS}
\end{figure}

The neutron star masses inferred for the 3:2 epicyclic resonance model by the considered EoS (Fig.~\ref{figure:MJEoS}a) are very low compared to the "canonical" value of $1.4M_{\sun}$.  For the non-rotating case, the implied NS configurations are in addition insufficiently compact to fulfill the modulation condition in Eq.~(\ref{equation:modulation-condition}). We find that this condition is satisfied only for high spin values, above $800$Hz, and strange matter EoS (MIT) (see the shaded region in Fig.~\ref{figure:MJEoS}b).

Searching through the region, we find the highest mass satisfying Eq.~(\ref{equation:modulation-condition}) to be $M=0.97M_{\sun}$. The related NS spin is 960Hz. This mass and spin correspond to $\nuL^{3:2}=580$Hz. For higher frequencies $\nuL^{3:2}$, the required mass is even lower. For $\nuL^{3:2}=630$Hz, it is $M=0.85M_{\sun}$, whereas the related NS spin is 900Hz.

For compact objects in the NS kHz QPO sources, there are at present no clear QPO independent mass estimates. In contrast, there is convincing evidence of the spin of several sources from the X-ray burst measurements \citep[see, e.g,][]{str-bil:2006}. In the group of sources discussed by \citet{abr-etal:2005a,abr-etal:2005b} considered in this paper, there are several that have spins in the range $\sim$250--650Hz. \emph{The NS parameters implied by the 3:2 epicyclic model therefore include not only very low masses, but also spins excluded by the QPO independent methods.}

The results obtained that falsify the epicyclic hypothesis are doubtless as far as:
\begin{itemize}
\item[i)] {The Paczy\'nski modulation mechanism is involved, implying that the inequality  \mbox{${R_{\mathrm{NS}}}<{r_{\mathrm{ms}}}$} is valid.}\\
\item[ii)] {The eigenfrequencies considered within the model are equal to (nearly) geodesic frequencies.}
\end{itemize}

The amplitudes of NS twin peak QPOs are often far higher than the BH amplitudes. The lensing effects are then insufficient for the observed modulation \citep[see e.g.,][]{bur-etal:2004,sch-rez:2006}. Nevertheless, there may be mechanisms other than Paczy\'nski modulation (that do not neccessarily require the \mbox{${R_{\mathrm{NS}}}<{r_{\mathrm{ms}}}$} condition). One example\footnote{\change{{The authors thank L. Rezzolla for suggesting this possibility during a discussion at the Relativistic Whirlwind conference in Trieste (June 2010).}}} may be an equilibrium torus without a cusp that oscillates with a high oscillation amplitude. In this case, accretion onto the neutron star may yet occur due to the overflow of the critical equipotential. The largest configurations fullfilling the 3:2 frequency condition for a non-rotating NS have the radius $1.8\!\times\!6M\!=\!10.8M$, which equals the radius of the 3:2 resonant orbit for $j=0$. Since all the considered EoS infer a NS radius below the resonant radius, the model should, in principle, work for this hypothetical mechanism. In addition, interference between a terminating disc and spinning NS surface near the 3:2 resonant orbit could represent a powerful excitation mechanism \citep[see also][]{lam-col:2003}. Nevertheless, some difficulties are apparent. In particular, the (unexplored) mechanism should be adjusted to the flow coming from the binary companion. {The twin QPO peaks with slowly varying centroid frequencies sometimes appear in the NS PDS for a few tens of minutes representing a timescale of $10^6$ oscillations. We note that there is an intrinsic instrumental fragmentation of observations that occurs on the same timescales and it is then assumed that the same QPO phenomenon often survive even longer \citep{Kli:2006:CompStelX-Ray:}. It would thus be neccesary to have an accretion flow with a neither low, nor high accretion rate that supports the considered torus-like configuration for a very long time.}

As quoted in Sect.~\ref{section:introduction}, for the 3:2 epicyclic resonance the values of the resonant eigenfrequencies for a non-geodesic flow are higher than those calculated for a nearly geodesic motion \citep{bla-etal:2007,str-sra:2009}. It has been shown that in a certain case the difference can reach about $15\%$.  
This would change the non-rotating mass to a higher value, $\sim1.2M_{\sun}$. 
From Fig.~\ref{figure:EoS} we can see that this value rather \emph{barely} fits the modulation condition for the MIT EoS. It thus cannot be fully excluded that the model is compatible with observations if the flow is fairly non-geodesic. 
However, needless to say that a serious treatment of this possibility will require investigation of the related pressure effects on the disc structure in the Hartle-Thorne geometry, since the aforementioned studies only consider Schwarzschild (in a pseudo-Newtonian approximation) and Kerr geometry. Moreover, they  assume a constant specific angular momentum distribution within the disc, while
there is evidence from numerical simulations of the evolution of accreting tori that real accretion flows tend to have rather near-Keplerian distributions \citep[e.g.][]{haw:2000,vil-haw:2003}. 
In this case, one expects the pressure corrections to be considerably smaller than those calculated for the marginal case of a constant angular momentum torus. However, there is no clear guarantee of this expectation and further investigation will be neccessary to resolve this issue.

\emph{We can conclude that the resonance model for NS kHz QPOs should involve a combination of disc-oscillation modes that differ from the geodesic radial and vertical epicyclic modes, or a modulation mechanism that differs from the Paczy\'nski modulation. The results also suggest that the two modes together with the considered modulation may operate as long as a fairly non-geodesic accretion flow is assumed for strange- or some nuclear- matter EoS.}

\section*{Acknowledgments}
\change{This work has developed from several debates initiated and richly contributed to by Marek Abramowicz and Wlodek Klu\'{z}niak in the past few years. We appreciate useful suggestions and comments of an anonymous referee, which helped to improve it. The paper has been supported by the Czech grants MSM~4781305903, LC~06014, and GA\v{C}R 202/09/0772. The authors also acknowledge the internal student grant of the Silesian University in Opava, SGS/1/2010. Part of the work reported here was carried out during the stay of GT and ZS at the G\"oteborg University, co-supported by The Swedish Research Council grant (VR) to M. Abramowicz.}


\end{document}